\documentclass[aps,twocolumn,groupedaddress,showpacs]{revtex4}
\usepackage{graphicx}
\begin{document}
\bibliographystyle{apsrev}


\title{Fermi Surface Evolution Across Multiple Charge Density Wave Transitions in ErTe$_{3}$}


\author{R. G. Moore$^{1,2}$}
\author{V. Brouet$^{3}$}
\author{R. He$^{1,2}$}
\author{D. H. Lu$^{1}$}
\author{N. Ru$^{2}$}
\author{J. -H. Chu$^{2}$}
\author{I. R. Fisher$^{2}$}
\author{Z. -X. Shen$^{1,2}$}

\affiliation{$^1$Stanford Synchrotron Radiation Laboratory, Stanford Linear Accelerator Center, Menlo Park, CA
94025, USA}

\affiliation{$^2$Geballe Laboratory for Advanced Materials and Department of Applied Physics, Stanford University
Stanford, CA  94305, USA}

\affiliation{$^3$Laboratoire de Physique des Solides, Universit$\acute{e}$ Paris-Sud, B$\hat{a}$t 510, UMR 8502,
91405 Orsay, France}



\date{\today}

\begin{abstract}
The Fermi surface (FS) of ErTe$_{3}$ is investigated using angle-resolved photoemission spectroscopy (ARPES). Low
temperature measurements reveal two incommensurate charge density wave (CDW) gaps created by perpendicular FS
nesting vectors. A large $\Delta_{1} = 175$ meV gap arising from a CDW with $c^{*}-q_{CDW1}\sim 0.70(0) c^{*}$ is
in good agreement with the expected value. A second, smaller $\Delta_{2} = 50$ meV gap is due to a second CDW with
$a^{*}-q_{CDW2} \sim 0.68(5) a^{*}$. The temperature dependence of the FS, the two gaps and possible interaction
between the CDWs are examined.
\end{abstract}

\maketitle

Charge density wave systems have been studied for many decades~\cite{gruner_94}.  CDWs and spin density waves, the
spin analog, exist in all dimensions.  However, enhanced electron interactions in systems with reduced
dimensionality increases the susceptibility for CDW formation, also known as a Peierls distortion in one
dimension~\cite{grioni_jes99,dardel_prl91,perfetti_prb02,fawcett_rmp88,schafer_prl03,gweon_prl98,brouet_prl04}. In
ideal 1D systems the FS topology determines the susceptibility for CDW formation via electron-phonon
coupling~\cite{gruner_94}. If the FS can be nested with one q-vector of a particular phonon mode, the ground state
energy can be reduced by electron-phonon coupling resulting in gaps opening at the Fermi level. Real systems are
never perfectly 1D, resulting in imperfect FS nesting with a partially gapped FS and residual metallic
pockets~\cite{schafer_prl03,gweon_prl98,brouet_prl04}. While a great wealth of information has been learned from
CDW systems, microscopic models explaining CDW mechanisms have been elusive and
debated~\cite{yao_prb06,johannes_prb06,johannes_prb08,malliakas_jacs05}.  It has even been proposed that the idea
of FS nesting is insufficient to explain the nesting vectors realized in the family of rare earth tritelluride
(RTe$_{3}$) compounds~\cite{johannes_prb08}. The RTe$_{3}$ family offers a unique opportunity to systematically
study CDW formation over a wide range of tunable parameters via rare earth
substitution~\cite{dimasi_prb95,ru_prb06,ru_prb08,brouet_prb08,malliakas_jacs05}. In addition, the simple
electronic structure makes RTe$_{3}$ more tractable for theoretical modeling to gain a deeper understanding of the
CDW phenomena and electron correlations in general~\cite{brouet_prl04,yao_prb06,laverock_prb05}.

\begin{figure}
\includegraphics[keepaspectratio=true, width = 3.4 in] {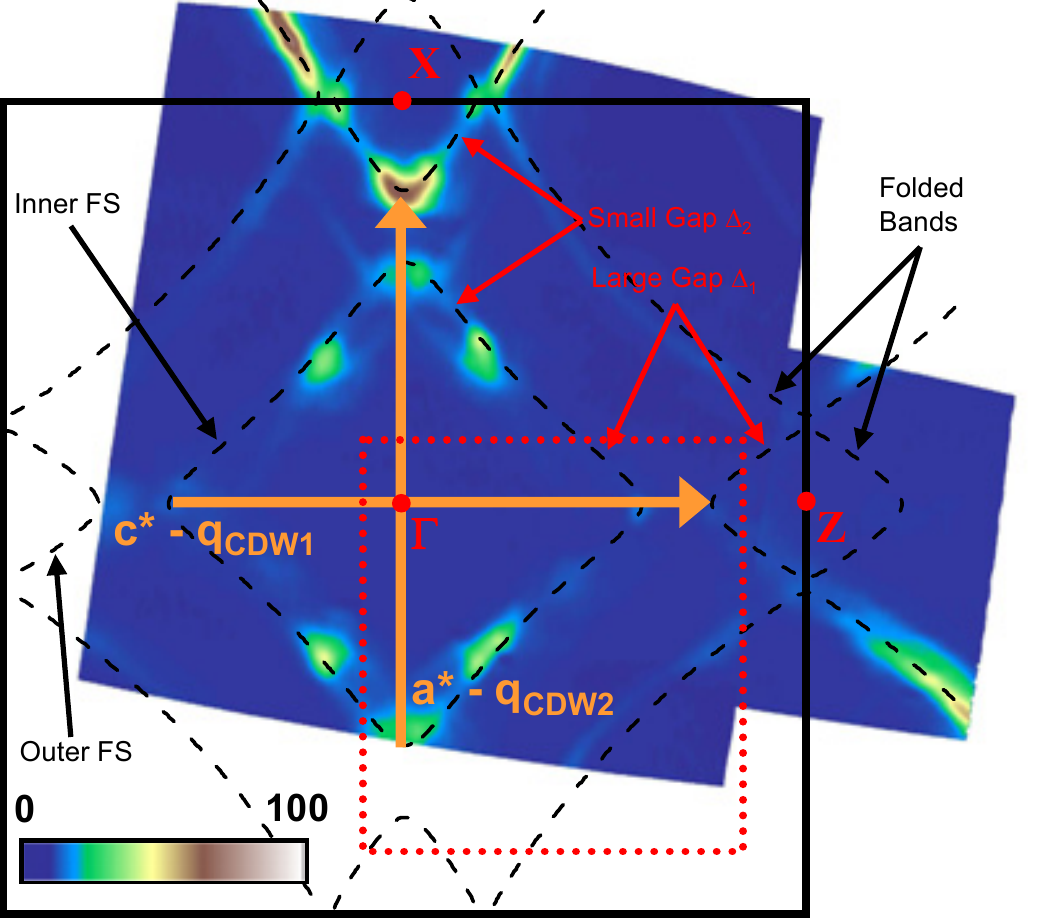}
\caption{ErTe$_{3}$ FS determined from ARPES.  Data taken at $T = 10$~K with $h\nu = 23.5$~eV and a $10$~meV
integration window around $E_{F}$. Solid square outlines $1^{st}$ BZ of the true 3D crystal structure.  Dashed
lines are expected FS from 2D Te net unmodulated by CDWs.  Red dotted square outlines FS areas shown in
Fig.~\ref{GapVST}.} \label{FS}
\end{figure}

The CDW properties for several members of the RTe$_{3}$ family have been well characterized
previously~\cite{gweon_prl98,brouet_prl04,ru_prb06,ru_prb08,brouet_prb08,dimasi_prb95}. RTe$_{3}$ has a layered,
weakly orthorhombic crystal structure (space group $Cmcm$) consisting of two planar Te nets sandwiched between
buckled RTe slabs. X-ray analysis reveals an incommensurate lattice modulation characterized by a single wave
vector ($q \sim 2/7 c^{*}$)~\cite{ru_prb06,ru_prb08}. Previous ARPES studies show a partially gapped FS with the
gap maximum occurring along the $c$-axis (i.e. $\Gamma - Z$ or $k_{x} = 0$ since the $b$-axis is the long
axis)~\cite{gweon_prl98,brouet_prl04,brouet_prb08,komoda_prb04}. The gap evolution as a function of R is
consistent with a FS nesting driven sinusoidal CDW with the bandwidth and density of states at the Fermi level
($N(E_{F})$) tuned by a shrinking unit cell~\cite{ru_prb08,brouet_prb08,laverock_prb05}.

The recent discovery of a second CDW transition in heavier members of RTe$_{3}$ is a rare display of the riches in
CDW formation and a unique opportunity to further test the concept of a FS driven CDW.  The second CDW was first
discovered by transport measurements~\cite{ru_prb08}.  It has further been confirmed by x-ray scattering
displaying a second lattice modulation perpendicular to the first, parallel to the $a$-axis with $q_{2} \sim
1/3a^{*}$~\cite{ru_prb08}. We have performed ARPES investigations of ErTe$_{3}$, specifically chosen since it has
two well separated CDW transitions at $155$~K and $267$~K. Our finding further strengthens the notions of FS
nesting driven CDW formation in this particular compound. Data suggest that as the rare earth ion is varied, the
second CDW is formed only when the first CDW weakens with the decreasing lattice parameter, making larger FS
segments available for the new nesting condition to form~\cite{ru_prb08,brouet_prb08}. While the result can be
qualitatively explained by a simple tight binding (TB) model, the data contain rich subtleties suggesting a
dynamic interplay of the two CDWs. Our finding establish ErTe$_{3}$ as an excellent model system to study the
evolution and entanglement of two many-body states existing within the same atomic plane.

Single crystals used in this study were grown by slow cooling a binary melt and have been well characterized
elsewhere~\cite{ru_prb08}. All ARPES data were taken at the Stanford Synchrotron Radiation Laboratory beamline 5-4
with an energy resolution of $10$~meV and angular resolution of $0.3^{\circ}$.

\begin{figure}
\includegraphics[keepaspectratio=true, width = 3.4 in] {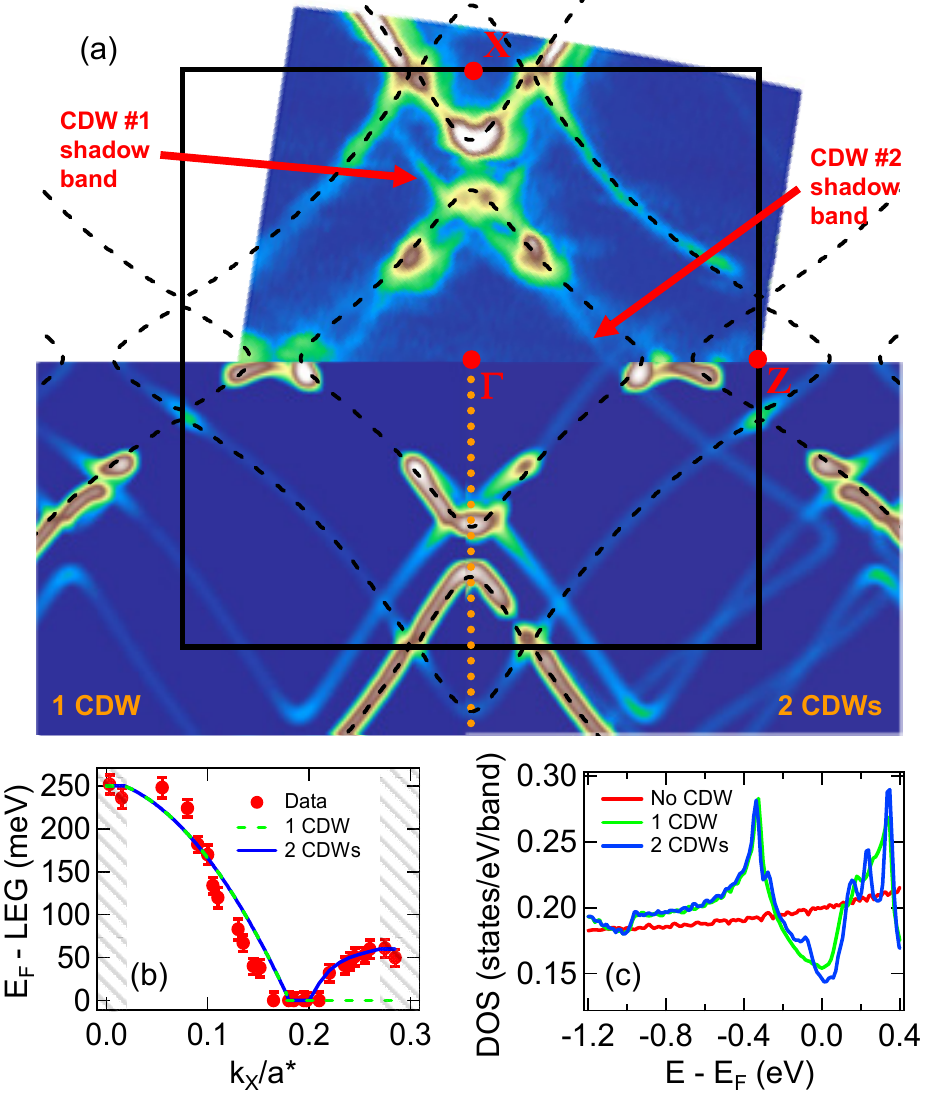}
\caption{(a) ARPES FS (top) compared to interacting TB model FS (bottom) with either one (left) or two (right)
CDWs. Both the data and model use a logarithmic intensity scale and $20$~meV integration window about $E_{F}$ to
emphasize weaker features just below $E_{F}$. The TB model was convoluted with an angular resolution function.
Vertical dotted line distinguishes between TB model with one or two CDWs.  (b) Energy of leading edge gap (LEG) in
the dispersion relative to $E_{F}$ measured along the inner FS for a constant $k_{z}$. Note the CDW gap magnitude
is measured relative to the center of the gap (see Fig.~\ref{GapVST}). The markers are the data and the lines are
from the TB model. Shaded areas correspond to regions where entire CDW gap is below $E_{F}$ resulting in intensity
at $E_{F}$ (see text). (c) Density of states from the TB model with different number of CDWs present.}
\label{TBModel}
\end{figure}

The FS determined from ARPES is shown in Fig.~\ref{FS}.  A large gapped region in the vicinity of the $\Gamma - Z$
axis is consistent with previous studies of RTe$_{3}$ members with a single CDW. While x-ray data shows $q_{CDW2}$
to be parallel to the $a$-axis, the FS along $\Gamma - X$ ($k_{z} = 0$) is not completely gapped at $T=10$~K as
shown in Fig.~\ref{FS}. The intensity is weak along the inner FS, but the outer FS retains significant spectral
weight. Further inspection slightly off the $\Gamma-X$ axis ($k_{z} \sim 0.07 c^{*}$) reveals a second gap does
appear in both the inner and outer FS pieces. In addition, several weaker features are evident in the regions
around the large and small gaps just below $E_{F}$. Fig.~\ref{TBModel}a is of the same data as Fig.~\ref{FS}, but
with a logarithmic intensity scale and slightly larger integration window to emphasize the weaker features near
$E_{F}$. As shown later, the weaker features provide crucial evidence for the induced FS from both CDWs. Given the
complexity, there is amazing degree of agreement between data and the expected FS pieces.

While ARPES exposes distinct differences between the FS of lighter RTe$_{3}$ members with a single CDW and the FS
of ErTe$_{3}$~\cite{gweon_prl98, brouet_prl04,brouet_prb08}, differences in the theoretical FS are
subtle~\cite{ru_prb08}. Linear muffin tin orbital (LMTO) calculations suggest the electronic structure near
$E_{F}$ is still dominated by the $p_{x}$ and $p_{z}$ orbitals of the two Te
planes~\cite{laverock_prb05,ru_prb08,laverock_private}. Hence, one would expect that the TB model of a single Te
plane which so accurately describes other rare earth members would also accurately describe ErTe$_{3}$. Details of
the TB model are described elsewhere~\cite{brouet_prl04,brouet_prb08}. A square net of Te atoms is assumed with
perpendicular chains created by the in-plane $p_{x}$ and $p_{z}$ orbitals ($p_{y}$ orbitals are assumed to be
filled).  The model parameters consist of electron hopping terms along a particular chain ($t_{\parallel}$) and
perpendicular to the chain ($t_{\perp}$).  The Fermi velocity, the slope of the bands traversing $E_{F}$, is
primarily determined by $t_{\parallel}$ while a FS curvature is introduced proportional to
$t_{\perp}$/$t_{\parallel}$. In this work the TB model is expanded to not only include effects of the $\sqrt{2}
\times \sqrt{2}$ $R45^{\circ}$ reduced 3D Brillouin zone on a 2D Te plane and $\pm(c^{*}-q_{CDW1})$ as used
previously, but includes the second CDW with the addition of $p_{x}$/$p_{z}$ bands translated by
$\pm(a^{*}-q_{CDW2})$.

To further understand the implications of two transverse CDWs on the FS, the TB model is further extended to
include interactions between various bands.  An electron-phonon coupled Hamiltonian is used to model the band
interactions in a perturbative fashion~\cite{gruner_94,brouet_prb08}.  Interactions arising due to the 3D crystal
structure and both CDWs are included in the Hamiltonian, however, only first order interactions due to band
crossings arising from a single $q$-vector translation are included.  While higher order terms are expected to
exist due to the incommensurate nature of the CDWs, the intensity is expected to be extremely weak and thus can be
neglected~\cite{brouet_prl04,brouet_prb08,voit_sci00}. The lattice parameters ($a = c = 4.27$\AA$ $ assuming a
square Te net) are taken from x-ray results and all other model parameters are estimated from the data, both from
the FS and band crossings occurring below $E_{F}$.  Since the intensity appearing in the reconstructed FS
originates from the original $p_{x}$ and $p_{z}$ bare bands, the color scale in the model FS is proportional to
the sum of the square of the resulting $p_{x}$ and $p_{z}$ eigenvector amplitudes from the interacting
Hamiltonian~\cite{brouet_prb08}. The resulting reconstructed FS from the interacting TB model is shown in
Fig.~\ref{TBModel}a. The spectral intensity existing near the FS corners at both $k_{x}=0$ and $k_{z}=0$ are now
explained by the electron-phonon coupled TB model of the CDW.  Shadow bands, bare $p_{x}$ and $p_{z}$ bands
translated by CDW $q$-vectors, correspond to weak features observed in the data revealing the two transverse CDWs.

Due to the agreement between the model FS and the observed one, the extended interacting TB model is used to
characterize the CDW properties in a similar manner to previous studies of lighter RTe$_{3}$
members~\cite{brouet_prb08}. The gap evolution along $k_{x}$ for the inner FS square is shown in
Fig.~\ref{TBModel}b, demonstrating the effects of multiple CDWs.  The excellent agreement between the model and
experimental data suggests the observed FS is consistent with a nesting driven CDW, however, quantitative
discrepancies exist which must be discussed. The model intensity at the corners of the inner FS square do not
match the experimental data. This discrepancy is observed both along the $\Gamma - X$ and $\Gamma - Z$ directions,
but is most pronounced along $\Gamma - Z$.  This discrepancy is most likely due to the simplicity of the model,
neglecting effects from higher order CDW terms, interactions beyond the nearest neighbor in-plane $p$-orbitals, a
bi-layer splitting arising from neighboring Te planes and orthorhombicity of the unit cell~\cite{brouet_prb08}.
Despite these discrepancies, the agreement justifies using such a simple model for illustration purposes. The
density of states is calculated for the interacting TB model and shown in Fig.~\ref{TBModel}c. The onset of a
single CDW suppresses $N(E_{F})$ to $\sim 77\%$ of the unmodulated value while the second CDW further suppresses
$N(E_{F})$ to $\sim 74\%$. Although the gains due to the second CDW are modest, mean field transition temperatures
depend exponentially on $N(E_{F})$. In addition, estimations of the Te bandwidth for GdTe$_{3}$ ($c=4.33$\AA$ $)
is $4.70$~eV~\cite{brouet_prb08} while for ErTe$_{3}$ ($c = 4.29$\AA$ $) it is $4.85$~eV. While ErTe$_{3}$ ARPES
data is the first in the RTe$_{3}$ family to show multiple gaps in the FS, the increasing bandwidth proportional
to the shrinking lattice agrees with the general RTe$_{3}$ trend.

\begin{figure}
\includegraphics[keepaspectratio=true, width = 3.4 in] {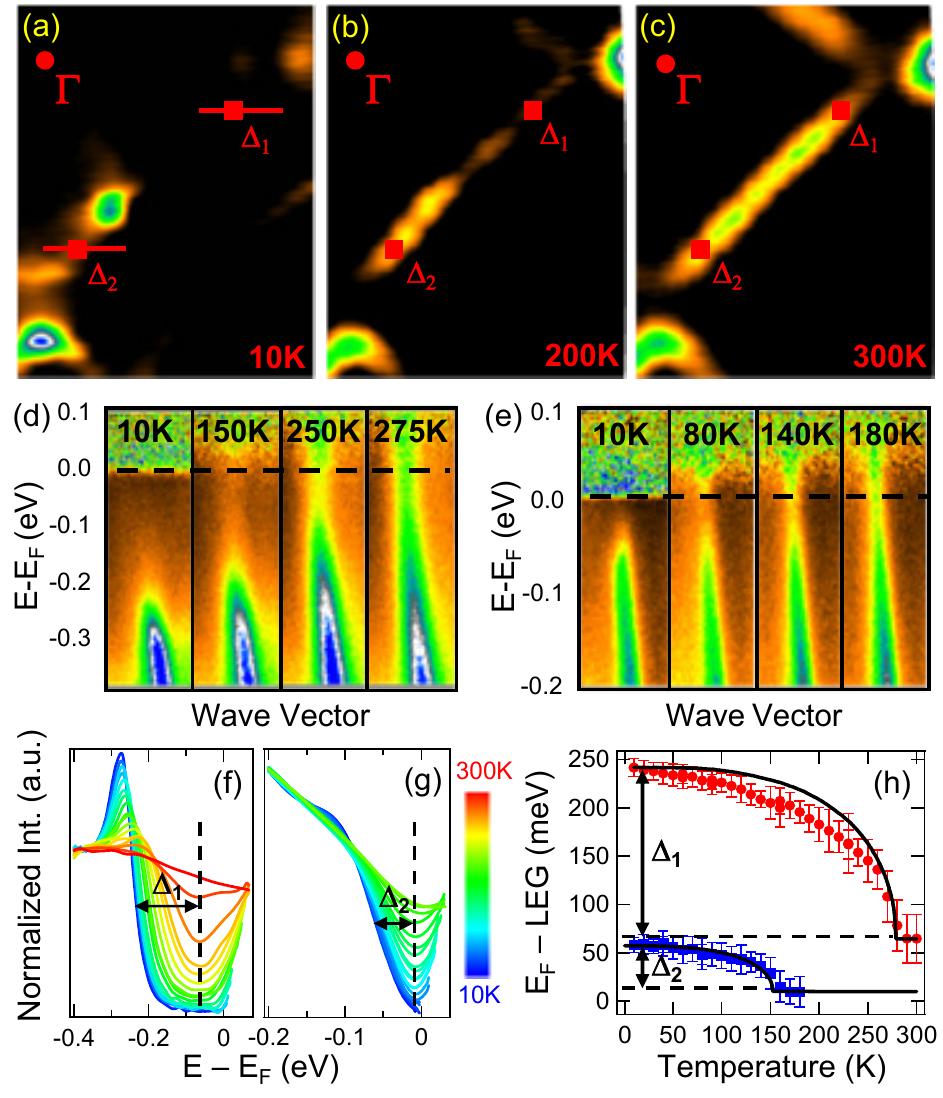}
\caption{(a-c) ErTe$_{3}$ Fermi surface taken at different temperatures.  The data shows one section of the inner
FS to emphasize the effects of the closing gaps.  Markers correspond to locations of the measured gaps. Horizontal
lines in (a) represent extent of cuts shown in (d-e). (d-e) Spectra divided by the Fermi function at different
temperatures for the large and small gaps respectively. (f-g) Temperature dependence of the fitted and denoised
dispersion data for the large and small gaps respectively. Different colored lines correspond to different
temperatures.  Dashed vertical lines mark the center of the closing gaps. (h) Summary of the two energy gaps for
different temperatures. Solid lines show expected mean-field behavior with energy gaps and transition temperatures
scaled to experimental values.  Note that the magnitude of each gap is measured from the center of each gap marked
by horizontal dashed lines.} \label{GapVST}
\end{figure}

Studies of lighter RTe$_{3}$ members with a single CDW suggest a shrinking gap accompanies the shrinking lattice
parameter, allowing a larger portion of the FS to remain intact.  In addition, the gap maximum always appears
along the $c$-axis~\cite{brouet_prb08}. Perfectly nesting the corners of the FS along each axis (i.e.
$c^{*}-q_{CDW}\sim 0.68c^{*}$) will pin the center of the CDW gap at $E_{F}$~\cite{brouet_prb08}. However,
observed nesting vectors and hybridization of $p_{x}$ and $p_{z}$ bands effectively pushes the center of the gap
below $E_{F}$.  This effect is most pronounced at the corners of the inner and outer FS allowing intensity to be
observed in these regions.  The gap magnitudes ($\Delta_{1}=175$~meV and $\Delta_{2}=50$~meV) are smaller in
ErTe$_{3}$ than the lighter rare earth family members (for CeTe$_{3}$ $\Delta\sim 400$~meV~\cite{brouet_prl04})
and the observed FS curvature ($t_{\perp}/t_{\parallel}$) is $\sim 15\%$ greater in ErTe$_{3}$. These differences
account for the large intensity observed along $\Gamma-X$ as one might expect this region to be gapped at $E_{F}$
based on previous RTe$_{3}$ studies. With smaller gaps and larger FS curvature, the downward shift of the CDW gaps
due to the length of the CDW $q$-vectors and $p_{x} - p_{z}$ hybridization allows intensity from the top of the
CDW gaps to be observed at $E_{F}$.

To further illustrate the implications of CDW formation on the FS, temperature dependent FS data are taken.  One
significant advantage of ErTe$_{3}$ is the fact that the room temperature structure is free from any
CDW~\cite{ru_prb08}. Thus this system provides a unique opportunity to observe the FS evolution across multiple
CDW transitions. The large gaps observed at low temperatures for this family allow us to observe the gap opening
despite thermal broadening effects. Fig.~\ref{GapVST}a-c show the FS data taken at different temperatures for one
side of the inner FS and Fig.~\ref{GapVST}d-g show the temperature dependence of $p_{x}/p_{z}$ bands near the
large and small gaps with a CDW and without. At $T = 200$~K only one gap is evident, while at $T = 300$~K the
inner FS square is fully closed as expected from a structure unmodulated by any CDW~\cite{laverock_prb05}.  To
track the temperature evolution of the leading edge gap, the spectra were first divided by the temperature
dependent Fermi-Dirac function convoluted with an energy resolution function~\cite{lee_nat07}. Such a procedure
allows the determination of the center of the energy gap ensuring accurate measurements of gap values. In
addition, since the center of the energy gap is observed below $E_{F}$, CDW $q$-vectors used in the TB model can
be determined directly from the data resulting in $c^{*}-q_{CDW1} \sim 0.70(0)c*$ and $a^{*}-q_{CDW2} \sim
0.68(5)a*$. The CDW wave vectors determined from ARPES are in excellent agreement with the lattice modulation
vectors observed in x-ray data~\cite{ru_prb08}.  The leading edge gap was then determined by fitting momentum
distribution curve peaks with Lorentz functions and tracking the point of inflection in the fitted band dispersion
after denoising via wavelet shrinkage. Instead of tracking the gaps at $k_{x}=0$ and $k_{z}=0$, temperature
dependent data were taken at $k$-points where the gap maxima with no FS intensity are observed. Fig.~\ref{GapVST}h
summarizes the temperature dependent data showing both gaps closing. A mean-field order parameter curve scaled to
the maximum observed gap is also plotted for comparison. The smaller $\Delta_{2}$ and larger $\Delta_{1}$ gaps are
observed to close at $T_{C2}\sim 160$~K and $T_{C1}\sim280$~K, respectively, in good agreement with the transport
and x-ray data~\cite{ru_prb08}. The development of the gaps appears to be second order within the experimental
uncertainty as no hysteresis has been observed.  While the closing of the gaps is suggestive of a mean-field type
behavior, $\Delta_{1} (T)$ is somewhat suppressed from the mean-field curve.  In addition, it should be noted that
$2\Delta_{1} / k_{B}T_{C1}$ is $\sim 2 ( 2\Delta_{2} / k_{B}T_{C2})$ while the area of the FS gapped by
$\Delta_{1}$ is $\sim 3$ times the area gapped by $\Delta_{2}$.


The observed $q$-vectors, the observation of CDW gaps below the Fermi-level and the use of an electron-phonon
coupled model Hamiltonian may suggest the FS plays little role in the formation of the CDWs~\cite{johannes_prb08}.
However, caution is advised as the model FS in Fig.~\ref{TBModel}a is not the one electron eigenvalues resulting
from the mixing of the different bands. To accurately model the FS, the calculated spectral weight ($p_{x}$ and
$p_{z}$ eigenvectors) had to be used. The FS in Fig.~\ref{GapVST}c is also a poor match for the model FS
eigenvalues with no CDWs because the bare bands folded back into the reduced 3D Brillouin zone are too weak to be
observed. Lowering of the ground state energy is achieved by gapping the FS and the model spectral weight suggest
the shape of the FS could still play a significant role in the CDW formation.

Both CDWs exist within the same Te plane~\cite{ru_prb08}, thus both CDWs modulate the positions of the same Te
atoms. Hence, ErTe$_{3}$ offers a unique opportunity to directly study the crossover from quasi-1D to quasi-2D
behavior. Upon initial inspection, each CDW appears uni-directional and completely decoupled. However, suppression
of $\Delta_{1}(T)$ from the mean-field curve, the discrepancy between $2\Delta_{1}/k_{B}T_{C2}$ and
$2\Delta_{2}/k_{B}T_{C2}$ still need to be explained. Such discrepancies may arise due to the interplay between
the two many body states. Subtle complexities arising from the crystal structure could interfere with the delicate
balance between the lattice and electronic energies, allowing for interactions between the two CDWs to arise. More
experimental and theoretical work is required to explore such possibilities.

\begin{acknowledgments}
We thank S. Kivelson, H. Yao, E. -A. Kim, J. Laverock and S. B. Dugdale for insightful discussions regarding our
data and model.  SSRL is operated by the DOE Office of Basic Energy Science, Division of Chemical Science and
Material Science. This work is supported by DOE Office of Science, Division of Materials Sciences, with contract
DE-FG03-01ER45929-A001 and DE-AC02-76SF00515.
\end{acknowledgments}


\bibliography{Moore_ErTe3_PRL}

%
%

%
%

\end{document}